\documentclass{sig-alternate-10pt}
\usepackage{notoccite}
\usepackage[numbers,sort&compress]{natbib}
\usepackage{graphicx}
\usepackage{times}
\usepackage{xspace}
\usepackage[svgnames]{xcolor}
\usepackage[english]{babel}
\usepackage{multirow}
\usepackage{booktabs}
\usepackage{array}
\usepackage{comment}
\usepackage{soul}
\usepackage{algorithm}
\usepackage{algorithmic}
\newcommand{\specialcell}[2][c]{%
\begin{tabular}[#1]{@{}c@{}}#2\end{tabular}
}

\usepackage{float}

\newcolumntype{P}[1]{>{\centering\arraybackslash}p{#1}}

\usepackage[hyphens]{url}
\usepackage{hyperref}
\hypersetup{breaklinks=true}

\usepackage{tikz}
\usepackage{graphicx}
\usepackage{url}
\usepackage{thumbpdf}
\usepackage{amsmath}
\usepackage{listings}
\usepackage{balance}
\usepackage{clrscode}
\usepackage{enumitem}
\usepackage{gensymb}
\usepackage[utf8]{inputenc}
\usepackage{amsthm}
\theoremstyle{plain}

\usepackage{dsfont,bm}

\newcolumntype{v}[1]{>{\centering\let\newline\\\arraybackslash\hspace{0pt}}p{#1}}

{\end{list}}
\makeatletter
\def\BState{\State\hskip-\ALG@thistlm}
\makeatother

\usepackage{listings}
\numberofauthors{5}
\date{}

\title{Towards A Domain-Customized Automated Machine Learning Framework For Networks and Systems}
\author{Behnaz Arzani, Bita Rouhani \\ \textit{ ``We are drowning in information and starving for knowledge'' -- John Naisbitt.}}

\begin{document}

\sloppy

\maketitle

\textbf{Abstract --} Clouds gather a vast volume of telemetry from their networked systems which contain valuable information that can help solve many of the problems that continue to plague them. However, it is hard to extract useful information from such raw data. Machine Learning (ML) models are useful tools that enable operators to either leverage this data to solve such problems or develop intuition about whether/how they can be solved. Building practical ML models is time-consuming and requires experts in both ML and networked systems to tailor the model to the system/network (a.k.a ``domain-customize'' it). The number of applications we deploy exacerbates the problem. The speed with which our systems evolve and with which new monitoring systems are deployed (deprecated) means these models often need to be adapted to keep up. Today, the lack of individuals with both sets of expertise is becoming one of the bottlenecks for adopting ML in cloud operations. This paper argues it is possible to build a domain-customized automated ML framework for networked systems that can help save valuable operator time and effort. 

%ML shows promising results in distill information, extract useful extraction. As such, it is expected to enable automated large scale data analysis...

\vspace{-2mm}
\section{introduction}

%There has been a rise in using Machine Learning (ML) to solve problems in networking/systems~\cite{mao2017neural,arzani2016taking,winstein2013tcp,valadarsky2017learning,cortez2017resource,zhang2018deepview,mao2016resource,jay2018internet}. 

Cloud operators gather large volumes of telemetry from their networks and systems~\cite{arzani2016taking,roy2018cloud,roy2017passive, guo2015pingmesh,cortez2017resource,tammana2016simplifying,khandelwal2018confluo,yaseen2018synchronized}. While this data contains valuable information, it is hard for operators to extract insights or action-items from such large quantities of (high-dimensional) data~\cite{davidblei}. Such insights help solve challenging problems that continue to plague their networks and systems.  Today, operators use heuristics (often approximations of integer linear programs) and manually encoded rules to solve such problems~\cite{dhamdhere2007netdiagnoser,jalaparti2016dynamic,chen2014msplayer,arzani2014deconstructing,zhao2019minimal,holterbach2019blink,ghorbani2017drill,dharmapurikar2016weighted}. Such heuristics are (often clumsy) proxies for detecting patterns and making generalizations about networked systems. ML is a promising alternative (compliment) -- it can adapt over time (through retraining) to new conditions, can leverage historical data to make optimized decisions based on past observations, and learns complex properties of the problem~\cite{beluch2018power,bottolo2010evolutionary,lecun2015deep} e.g, ML has enabled large improvements in time-to-detection of cloud incidents~\cite{zhang2018deepview}. ML can help operators test their conjectures about whether a dataset can help solve a given problem. The insights derived from using ML can expose valuable, hidden, information that can help build solutions to such problems~\cite{arzani2016taking} even if the model is not used in production.

\noindent \textbf{Need.} While ML is useful, it does have several short-comings: sensitivity to the input feature-set, debugability, etc. Thus, building ML models for solving problems in networked systems often requires careful domain customization~\cite{arzani2016taking, mao2019park, mao2017neural}, to compensate for some of these short-comings.  Today, building such models requires someone with expertise in both domains (networking/systems and ML) to select the model, engineer features, identify corner cases where the model can fail, and optimize the model for performance and scale~\cite{mao2017neural,arzani2016taking,winstein2013tcp,valadarsky2017learning,cortez2017resource,zhang2018deepview,mao2016resource,jay2018internet,davemaltz}. Such resources are scarce and the cost, in time and human resources, of tackling such problems is high. This is because of the scale, diversity, and complexity of the data we collect~\cite{arzani2016taking,roy2018cloud,roy2017passive,arzani2018007,zhuo2017understanding}, the pace at which our networks and systems evolve, and the number of questions we want answered. Not all models are re-usable: every system is different and so a model built for one system may not easily translate to another.  Our conversations with operators of large public clouds indicate the lack of operators with joint expertise is the main bottleneck when trying to build ML models to solve problems in networked systems. Others have also observed similar challenges~\cite{davemaltz}.

\noindent \textbf{Proposal.} This paper argues we can automate the process of building ML models for cloud operators; but, such an automated framework must be {\em domain-customized}. Operators should not need to customize ML models for each problem, but they should be able to express high-level objectives to an automation framework that could then generate appropriate solutions. We need to build a framework that enables users to use ML to solve problems in networked systems without having in-depth ML expertise and that, similarly, enables ML experts to contribute to solving problems in networked systems without having expertise in these domains. Such a framework can also help operators to take a more principled approach to building ML models for production networks and systems. We assume operators have thought about whether ML-based solutions are appropriate for the problem they are solving before using the system. How to make this decision is itself a topic for research~\cite{davemaltz}.

\noindent \textbf{Promise.} The ML community has shown it is possible to build Automated ML (AutoML) frameworks. Such frameworks remove humans from the process of building ML pipelines. They have shown on-par performance, compared with human experts, when solving several problems~\cite{guyon2019analysis}. However, they aim to solve all problems posed to the system using the same underlying process~\cite{autosklearn,TPoT}. Such generic approaches will not work well for all classes of problems (see~\S\ref{sec:motivation})~\cite{liu2018very}. We have seen human-designed, domain-customized models yield high accuracy in video streaming~\cite{mao2017neural} (reinforcement learning), traffic engineering~\cite{valadarsky2017learning} (deep neural networks), diagnosis~\cite{arzani2016taking} (random forests), and other problems in networked systems. Park \cite{mao2019park} showed how to domain customize reinforcement learning models by restricting their exploration space based on the properties of the networking problem they are solving. We can build AutoML frameworks that can also generate such domain-customized models, on-par with those designed by human experts, faster, and at reduced cost.

Users can provide, as input, what they know about the problem, and the framework can use this context to generate domain-customized models. Such context can help constrain the search space, remove irrelevant features in the input data, and improve feature engineering so we can capture the most useful information about the problem in the model~(see~\S\ref{sec:motivation}).

\noindent \textbf{Contributions.} We formally define the domain-customized AutoML problem and show preliminary results demonstrating domain-customizing existing state-of-the-art AutoML systems can help improve their accuracy and performance. We explain why domain-customizing AutoML is not always straightforward, propose an architecture for a domain-customized AutoML framework, and outline open questions we, as a community, need to answer to realize it.

%%\TODO{Would be cool if we actually tried it and showed the results suck.}

\section{An Example in Regression}
\label{sec:motivation}
\vspace{-1mm}

We first show an example of adding context to AutoML. We look at a {\em straightforward} regression problem: predicting the expected latency of a VM to other VMs in its VNet given where its located.
A VNet is a virtual network between VMs in the same subscription. A potential use-case for this problem is network-aware VM placement~\cite{cortez2017resource}.

\noindent{\textbf{The dataset.}} We use data from a system similar to VNet Pingmesh~\cite{roy2018cloud} that monitors the health of the network between the VMs in each VNet. For each VM in the VNet, it records the average latency from that VM to other VMs in the VNet every minute (by sending pings to those VMs from the host). The raw data consists of the cluster the VM is in, it's VNet ID, the host on which it is deployed, the VM name, and the average latency to other VM's in the VNet. We use $2$ hours of VNet Pingmesh data from 1000 production clusters of a public cloud. We create a ``small'' and ``large'' subsets from this data for our experiments. The ``small'' subset uses the first $45$ minutes of data for training and the next $25$ minutes for testing. The ``large'' subset uses the first hour of data for training and the next $45$ minutes for testing. 

\noindent{\textbf{Evaluation metric.}} We use the coefficient of determination or $R^2$ score for our evaluation. The $R^2$ score is a value, $\le 1$, where the best possible score is 1. A constant model that always predicts the expected value of y, disregarding the input features, would have an $R^2$ score of 0.0. Negative scores indicate a model with worst accuracy than one that predicts the mean of the training data at all times.

\noindent{\textbf{The algorithm.}} We use an, open-source, state-of-the-art AutoML framework: Auto-Sklearn~\cite{autosklearn}. Auto-SKlearn is built on top of the popular ML toolkit SKlearn~\cite{sklearn}. It was the winner of the ChaLearn AutoML challenge~\cite{guyon2016brief}.

\begin{table*}[t]

\centering
{\scriptsize
%|l||p{16.2cm}|

    \begin{tabular*}{1\textwidth}{p{1cm} P{1.1cm}  P{1.1cm}  P{2cm} P{1.5cm} P{1.2cm} P{1.2cm} P{1.9cm} P{1.9cm} P{1.2cm}}
    \hline
       \textbf{Exp ID} &  
       \textbf{\specialcell{(1)}} & \textbf{\specialcell{(2)}} & 
       \textbf{\specialcell{(3)}} & \textbf{\specialcell{(4)}} & \textbf{\specialcell{(5)}} & \textbf{\specialcell{(6)}} & \textbf{\specialcell{(7)}} & \textbf{\specialcell{(8)}} & \textbf{\specialcell{(9)}}  \\
    \hline \hline
    \textbf{Description} & 
    Small training set & Larger training set & 
    Exp. (1) + test set partitioning & Normalizing data per cluster & 
    Training per cluster & 
    Removing VM name & 
    Adding \#VMs  in VNet (new feature) &
    Adding hosts in VNet (new feature) & 
    Exp. (8) + larger training set \\
   
    \textbf{$R^2$ score}  &  
    0.28$\pm$0.15 & 
    0.03$\pm$0.05 & -2.04$e^{19}\pm$3.4$e^{19}$ & 0.03$\pm$0.005 & 
    0.74 $\pm$0.06 &  
    0.74$\pm$0.06 & 
    0.73$\pm$0.06 & 
    0.65$\pm$0.1 & 
    0.69$\pm$0.08\\ \hline
     \end{tabular*}
     }
   
         \caption{Predicting latency between VMs in a VNet using VNet-Pingmesh data as input and by domain-customizing these inputs to Auto-SKlearn. Results are averaged over $40$ runs (with $95\%$ confidence interval). Each new experiment contains the changes in the previous one (all except (2) and (9) are on the small dataset).}
    
    \label{tab:results}
\end{table*}

\noindent{\textbf{Observations (Table~\ref{tab:results}).}} The raw data produces a low score (1). An ML expert may increase the amount of training data to improve accuracy, but the score doesn't improve (2). However, we know each cluster has different workloads and a different latency baseline: comparing the Pingmesh~\cite{guo2015pingmesh} latency of two clusters, the min -- and max, median -- latency in one is 4$\times$ that of the other\footnote{Note this is Pignmesh latency, not VNet Pingmesh latency.}. Hence, we use AutoML to find a different model for each cluster and achieve a $\sim$3$\times$ improvement in score (5). We also know many VM names are, typically, un-informative and so we remove them from the input, reducing the size of the training set and speeding up training without any loss in score (6).

Adding context is not always so simple. For example, we may decide to add where (which hosts) the other VMs in the VNet are deployed as a feature. This increases the number of features: as a rule of thumb, to maintain accuracy, as the number of features increases the number of training samples must also increase~\cite{bissmark2017sparse}. Indeed, using the same number of training samples resulted in a lower score (8). 

Skeptics may ask if it is possible that accumulation of error across clusters in Exp. (1)-(2) would explain these differences? We repeat Exp. (1) but partition the test set by cluster and average the score (Exp. (3)). Comparing Exp. (3) with Exp. (5)-(9) our conclusions remain unchanged. 

%These experiments show that expertise in either ML or networking on their own are insufficient to design the best ML model to solve this problem but there is need for expertise in both domains.

Our example is simple enough that operators could adjust their inputs themselves to fix the problem. Other problems may not be as simple. Consider~\cite{arzani2016taking} where the authors use TCP measurements along with a random forest-based model to classify whether a failure is caused by the client, server, or network. They found random forests to have poor accuracy ($\le 25\%$) when used unchanged. They leveraged the following insight to boost accuracy: because TCP is designed to detect most networking problems (and not others), it is possible to first, {\em accurately}, distinguish networking failures from other types of failures and remove them from the data and then proceed to classify other failures. This increased the information gain of individual features when identifying other classes of failures and improved overall accuracy. Finding such opportunities of improvement by an ML expert without networking background or, similarly, a networking expert without ML background is difficult and is what domain-customized AutoML aims to do.

\section{AutoML for Networks/Systems}
\vspace{-1mm}
We first describe the generic AutoML problem solved by state-of-the-art frameworks~\cite{autosklearn, TPoT} -- Combined Algorithm Selection and Hyper-parameter (CASH) optimization. We then discuss the limitations of such approaches and how we can address such challenges by carefully adding context.

\noindent \textbf{Definition 1} (CASH). Let ${A} = \{A_1, A_2,..., A_n\}$ be a set of algorithms with the corresponding hyper-parameter set of $\Gamma= \{\gamma_1, \gamma_2, ..., \gamma_n\}$. Further, let $L (A_i, \gamma_i, D_{train}, D_{test})$ denote the loss value obtained using algorithm $A_i$ with hyper-parameter set $\gamma_i$ that is trained with the training data $D_{train}$ and validated on the test data $D_{test}$. The goal of CASH optimization problem is to find the joint algorithm and hyper-parameter setting that minimizes:

\begin{align}
\vspace{-2mm}
A^*, \gamma^*  \in  \underset{A_i\in A, \gamma_i \in \Gamma}{argmin} ~ L (A_i, \gamma_i, D_{train}, D_{test}) \nonumber
\end{align}

Even though solving the generic CASH problem is a good starting point for automation, there is a significant gap between the achievable performance and the current state-of-the-art. This gap is mainly an artifact of the decoupling of CASH optimization from application context. The generic CASH problem suffers from (at least) three main limitations: 

\noindent{(1)} \textbf{Algorithm Space:} No single ML model outperforms all other models across {\em all} applications. Compiling a list of models and prioritizing them based on ``frequency of usage'' is not an optimal approach. For instance, the best models for solving networking problems may be different from those used to solve computer vision problems. Prioritizing algorithms based on their general popularity is sub-optimal. \\
(2) \textbf{Hyper-parameter Space:} Several ML methods (e.g., kernel-based SVM and regression) heavily rely on hyper-parameter optimization. Leveraging prior knowledge for a more effective hyper-parameter search can significantly reduce search time. This is particularly important for applications where users have a strict time/resource constraint. \\
(3) \textbf{Data Redundancy:} Redundancy in the input, both in the feature space and number of samples, impact both the quality of response and time to response. Distilling the data based on domain knowledge can significantly reduce noise in the input space and accelerate convergence. We discuss other ways context can help improve AutoML in section~\S\ref{sec:arch}.

We should reformulate the CASH problem so the objective captures context-aware AutoML frameworks that are simultaneously accurate, efficient, and easy to use. For a given application, the optimal algorithm/hyper-parameter space is a subset of the cross-product of different algorithm/hyper-parameter domains supported by the AutoML framework. This subset is often strict; meaning the active search space is conditional to the target application/constraint. 

More formally, the utility of an algorithm and its corresponding hyper-parameter domain is conditional to an application. Accounting for this property and conditioning algorithm, hyper-parameter, and data spaces by the domain results in a conditional tree-structured space (e.g., a directed acyclic graph) tailored to the target applications. Thus, the (hierarchical) CASH optimization problem is defined as:
%\begin{equation}
%\vspace{-1mm}
%\scalebox{0.9}{
%\begin{align}
%A^*, \gamma^*, f^* \in  &~\underset{A_i \in A}{argmin} ~ L^1 (A_i, D_{train}, D_{test}| S) \\
%&~\underset{f_i\in F, \gamma_i\in \Gamma}{argmin} ~ L^2 (\gamma_i, f_i(D_{train}), D_{test}| S, A_i). \nonumber
%\end{align}}
%\vspace{-1mm}
%\end{equation}

\begin{align}
A^*, \gamma^*, f^* \in  &~\underset{A_i \in A}{argmin} ~ L^1 (A_i, D_{train}, D_{test}| S) \\
&~\underset{f_i\in F, \gamma_i\in \Gamma}{argmin} ~ L^2 (\gamma_i, f_i(D_{train}), D_{test}| S, A_i). \nonumber
\end{align}

$S$ indicates the target application selected from a pre-defined set of applications supported by AutoML and $f_i(.)$s are transformation functions applied to the training data for better feature engineering. At each optimization step, AutoML aims to minimize the loss function over a carefully designed jointly dependent constrained set. The high-level algorithm is first selected to best fit the application and the input data. It then performs Hyper-parameter tuning and feature engineering conditional to the selected algorithm/application. In our tree-structured optimization, the choice made by a higher-level node impacts the choices available to lower-level nodes. Such nested hierarchical optimizations restrict the solution space and significantly reduce search time. They may even improve accuracy by reducing redundancy.

\vspace{-3mm}
\section{Design: Insights \& Challenges}
\label{sec:arch}
\begin{figure*}[t]
	\begin{center}
	\scalebox{0.91}{
		\vspace{-2mm}
		\includegraphics{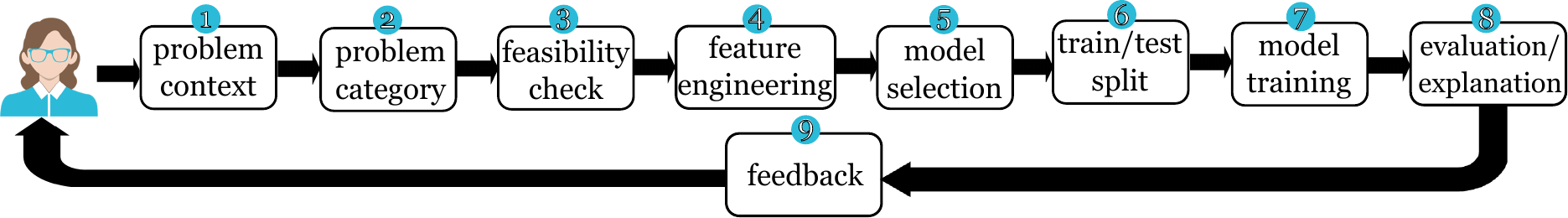}}
	\end{center}
	\vspace*{-0.8cm}
	\caption{\textbf{Our proposed architecture for a context-aware AutoML framework for networking.}}
	\vspace{-5mm}
	\label{fig:arch}
\end{figure*}

\vspace{-1mm}
Our design philosophy: we should not remove the human entirely from the loop but should leverage their domain knowledge without requiring them to have expertise in ML (Figure~\ref{fig:arch}). Each following subsection describes individual modules in the framework and the research questions that need to be answered to make them work in practice.

\vspace{-3mm}
\subsection{Providing problem context}
\vspace{-1mm}
The framework relies on context as an input. We expect users to input context through a domain specific language. The context can be high level -- the category of the problem being solved: congestion control, diagnosis, traffic engineering; or detailed -- e.g., trying to determine the right congestion window for TCP based on a specific congestion notification signal. Users may also choose to provide time and resource constraints which the framework should adhere to.

\noindent{\textbf{Open questions:}} What is the right set of abstractions the framework needs to expose to users? What is the desired level of detail we need so that we can generate accurate and performant (in terms of resource usage) ML models? How do we translate the input into the format required by the rest of the modules in the framework? How can we detect when we don't have enough information to find the right models?

\vspace{-3mm}
\subsection{Choosing problem category}
\label{subsec:category}
\vspace{-1mm}
There are many choices for what ML model to use: supervised vs unsupervised, regression vs classification, online vs offline learning. We need to decide which to use. Many existing AutoML frameworks expect users to make this determination themselves upfront~\cite{autosklearn,ghahramani2014automatic} but this may be difficult for users with limited to no knowledge of ML.

We can brute-force the set of available models to find the best fit or use Bayesian optimization or other algorithms to speed up the search~\cite{TPoT,autosklearn}. This approach can use-up a large portion of the available time/resource budget which takes away from other critical operations such as feature engineering, model evaluation, and hyper-parameter tuning.

\noindent{\textbf{Our insight:}} Context can help save resources when searching for  the right category -- knowing whether we are solving a latency estimation problem vs DDoS detection problem can help decide which models to use: we know labeled data will be scarce for DDoS detection and so we may decide to use (unsupervised) anomaly detection whereas we know that a regression algorithm is appropriate for latency estimation.

Auto-Sklearn~\cite{autosklearn} demonstrated  that meta-learning is effective in speeding up search. Meta-learning detects whether new datasets are similar to previously observed ones and starts the search with a model which was effective for those datasets in the past. Knowing the problem context can help improve meta-learning by reducing the number of datasets which are compared to the user's input and in determining which datasets are likely to be similar. Furthermore, it can also help create the right set of ``meta-features'' to use.

\noindent{\textbf{Open questions:}} To use context, we need to have a taxonomy of problems in networking, study which ML models are suitable for each, and find example data to use for meta-learning. For each class of problems in networking we need to identify ``similarity metrics'' to be used in meta-learning to identify whether two datasets in that domain are similar enough to warrant using the same model. For example, traffic engineering in the DC setting (either in the DC's wide area backbone or within the DC itself) may be a very different problem than traffic engineering in the wide area network~\cite{valadarsky2017learning}. We need a principled approach that helps identify metrics that can quantify whether these two problems are similar -- the best metric may depend on the problem being solved: congestion control vs video streaming.

\vspace{-3mm}
\subsection{Feasibility check}
\vspace{-1mm}
Many laymen in ML think it can solve any and all problems, but this is far from reality. It is essential to check whether the problem posed to AutoML is even solvable. Without a ``feasibility checker'', in practice, users may lose confidence in AutoML's ability and stop using it all-together when it fails to produce an accurate model. If AutoML can predict when and why it may fail, it can warn the user that the problem is too hard and tnat a human expert is needed.

We present an example from our own experience: Engineers of a production cloud wanted to use latency measurements from a Pingmesh~\cite{guo2015pingmesh}-like system to predict future packet drops and approached us for help. Configuration changes pushed by human operators can cause packet drops. Such changes are not captured in latency measurements (and no additional data is provided) -- it's unlikely for ML to produce high accuracy. To demonstrate this more concretely, we looked at the Spearman correlation~\cite{mukaka2012guide} between Pingmesh latency and packet discards observed on the top of the rack switches in two different DCs of that cloud: 0.029 and 0.064 are the results. Feeding this data into Auto-SKlearn yields an R2 score of $-0.027$, indicating poor accuracy. While these experiments are hardly proof that the problem is unsolvable, they indicate the problem is extremely difficult for the framework to solve and would benefit from expert analysis.

\noindent{\textbf{Our insight:}} Checking feasibility is hard. But context can help, at least, compute a difficulty measure for a given dataset. For example, the framework can (learn to) ask the operator a series of questions, given the context, which could help it compute whether it is likely to achieve acceptable accuracy. In the context of the example above, these questions may be: Is latency expected to be correlated with packet loss? Can humans influence the output being predicted? Does the input data capture human involvement? Are there external factors that the input is not capturing? 

We could also think about using context to tune statistical tests or other metrics~\cite{blumer1989learnability} to check feasibility. Context may help decide whether the results indicate we are likely to achieve acceptable accuracy for specific problem categories.

\noindent{\textbf{Open questions:}} How can we check feasibility? Statistical tests can help determine if the input and output are even correlated. A subset of the data could be used for an initial trial to determine the accuracy of the existing modules in the framework. Meta-learning may also prove useful.

It may be possible to build model-specific checks which would quantify whether it is likely for that model to produce acceptable accuracy. For example, one can check whether the input/output are linearly correlated before using linear regression. Such metrics can be used as a measure of feasibility for those models. Alternatively, VC dimensionality is a measure ML experts have developed to quantify the complexity of learning from a given input~\cite{blumer1989learnability}. We may be able to draw a mapping between the VC dimension of an input and the likelihood of AutoML's success. How useful these methods will be in practice is likely to depend on the category of problem we want to solve, and the solution may benefit from domain-customization.

One can also borrow from research in Human-Computer Interaction (HCI) to identify the right series of questions that, if posed to the user, can help determine the likelihood of solving network/systems related problem. An ML model (such as RL) may be even useful to determine, for each context, what series of questions are appropriate. 

Identifying the right metric from the above list (or maybe even finding a different one) is an open question.

\vspace{-3mm}
\subsection{Feature engineering}
\vspace{-1mm}

Prior work has shown good feature engineering can be crucial to building ML models with high accuracy~\cite{arzani2016taking}.

\noindent{\textbf{Our insight:}} Context helps infer the relationship between input features. Existing frameworks~\cite{kanter2015deep,CloudAutoML} exploit the structure in the input table to derive relationships across the columns of the input data. We can use context, and we can embed domain knowledge into the search algorithm itself to find the best features. For example, when using TCP statistics to find the entity responsible for a failure~\cite{arzani2016taking}, users can specify which features correspond to the source and destination IP. The framework can then create corresponding ``aggregate features'' that characterize normal and abnormal behavior for individual flows between those IP addresses over time. Once it arrives at a feature-set, the framework can then use auto-encoders~\cite{masci2011stacked} for dimensionality reduction.

\noindent{\textbf{Open questions:}} The problem type dictates the approach used for feature engineering: one may choose a different feature engineering scheme when designing a model for deciding the best bitrates to use for video streaming but opt for another when deciding the best congestion window for TCP. How to categorize these problems and what algorithms to use for feature engineering is an open problem.

\vspace{-3mm}
\subsection{Model selection}
\vspace{-1mm}

Many factors influence which model is the right choice for solving a given problem, e.g., the choice of model may depend on the resource and time constraints the operator specifies. Many models can be domain-customized through the choice of kernels -- independent variables can be pre-configured in the Covariance function of a Gaussian process; or the model structure -- encoding independence assumptions in a Bayes-Net. Other models can accept other ``priors''.

\noindent{\textbf{Our insight:}} Context can help identify the right priors to use for each model e.g., in many networking problems, topology and where the monitoring data is being collected can bear a huge impact on what priors may be appropriate (see~\S\ref{sec:motivation}).

\noindent{\textbf{Open problems:}} This may be one of the more challenging modules of the framework: when designing the DSL we need to anticipate what information may be useful to the framework for deciding upon the right priors, kernels, and models. Similarly, we need to decide how to interpret the user's input and use it for this purpose. The taxonomy of problems described in~\S\ref{subsec:category} can help answer these questions.  
\vspace{-3mm}
\subsection{Train/test split}
\vspace{-1mm}

Ensuring the right train/test split can be important when evaluating/comparing models. For example, take the problem of VM placement~\cite{cortez2017resource} -- we want to predict the resource usage of a VM given the past resource consumption of other VMs in the same subscription. 

If we split the data for train/test by time, data for individual VMs will be split between train/test sets, which can result in information leakage. The model will have high accuracy on the test set because it has ``seen'' the same VM in the training set. But as soon as it is used in practice, where it has to make predictions for new VMs, there will be a significant drop in accuracy. The solution is simple: ensure the data is split such that individual VMs are either in the training or the test set but not in both. The question is, how would an AutoML framework figure out that it needs to do this?

\noindent{\textbf{Our insight:}} Context can point us to potential causes of information leakage, e.g., in the previous example, knowing VMs should be considered separately. The framework can use the input context to derive what boundaries should not be crossed when splitting data into training and test sets.

\noindent{\textbf{Open questions:}} How to translate the information provided by the user into whether there is likely to be information leakage for a given train/test split? 

\vspace{-2mm}
\subsection{Evaluation/Explanation}
\vspace{-1mm}
Evaluation can help give feedback on how the user can modify their inputs (see~\S\ref{subsec:feedback}) to get higher accuracy and better models, as well as insight on how to interpret the results. 

\noindent{\textbf{Our insight:}} The right metric for evaluating a model may depend on the problem context. One can even use this metric when training and evaluating models such as reinforcement learning which need to optimize an expected reward. We can maintain a set of possible metrics, e.g., flow completion time (for congestion control design), buffer occupancy (for video streaming), link utilization (for traffic engineering), average peering costs (for traffic engineering), etc. The framework may even present a set of choices to the user and allow the user to decide which criterion is appropriate.

We can also translate ML-based evaluation metrics into human-readable text that explain the utility of the model. For example, the framework can use natural language processing to output sentences such as ``When the system \textit{outputs} <<the link has failed>> it is likely to be correct $80\%$ of the time'' where $80\%$ is the model's precision on the test set.

Explainable models also exist~\cite{lakkaraju2018human,wang2019designing,meineexplainable} and can help users understand what inputs have helped the model and also reduce data collection overheads -- operators can use these outputs to decide which measurements are helpful. There is often a tradeoff between accuracy and explainability. Users may decide on a preferred tradeoff which would determine whether AutoML favors explainable models.

Context can help provide more insights when explaining models and combined with natural language processing may even help improve the readability of the output.

\noindent{\textbf{Open questions:}} How would you use context to translate ML evaluation metrics such as F1 scores, precision/recall, accuracy, ROC curves, etc to metrics the operators care about? What are the right domain-specific performance metrics to use for each category of problem (e.g., for the reward function in reinforcement learning)? How do you derive those metrics from the input data? How do we use context to improve model explain-ability? 

\vspace{-2mm}
\subsection{Feedback}
\label{subsec:feedback}
\vspace{-1mm}

Humans should be kept in the loop when designing ML models for production systems. If AutoML can produce useful and human readable feedback then users may be able to adjust their inputs to build models with higher accuracy based on that feedback. Feedback can also help operators understand which inputs have helped the models and which one's haven't. They can use this information to, if necessary, reduce monitoring overhead in their networks and systems.

This module could take advantage of the problems the framework has solved in the past. Take the example where the framework was previously used to estimate the right TCP congestion window based on round trip times and bandwidth estimates and is now being used to do the same but with the bandwidth delay product as the input. The framework may observe that the previous input was more predictive of the correct output and provide this feedback to the user who can then adjust their inputs accordingly.

\noindent{\textbf{Our insight:}} AutoML can use historical records to inform the user if there are inputs that could improve the model's accuracy. Such feedback will depend on the problem context as, for example, a congestion control problem may not be directly comparable to a VM placement problem. 

\noindent{\textbf{Open questions:}} It is unclear what type of context can help improve the feedback the framework provides to the user. We also need to understand what this feedback should be and whether there is a \textit{confidence metric} we can assign to it. The \textit{confidence metric} can show the framework's confidence in the feedback and the user can use it to gauge what actions are appropriate based on that feedback. 

\vspace{-3mm}
\section{Discussion}
\vspace{-1mm}

\noindent{\textbf{Should ML be used to solve networking/systems problems?}} We focused our attention on the design of a domain-customized AutoML framework that enables networking/systems experts to build ML models for their networks/systems. Is ML the right solution for these problems? There has been much debate on the subject~\cite{davemaltz}. We believe there are cases where ML is a great tool: when the number of features describing the problem space is large and there is no first-principles understanding of the problem -- network availability problems are one such example; or when we are building our intuition about a problem before we solve it -- here, ML can help us develop our understanding about the problem and build better solutions. These arguments not-withstanding the community can benefit from a careful study of what types of problems can benefit from an ML-based solution and where ML is not the right choice. This, however, is beyond the scope of our AutoML framework.

\noindent{\textbf{On-line debugging of the models produced by AutoML.}} Our design (~\S\ref{sec:arch}) targets the steps involved in finding the right model to use to solve a particular problem using a given dataset. But, if we want to use these models in practice, we also need to track their performance in real-time and to replace them with new models -- either the same model re-trained or a completely different model -- when they start to become outdated. Thus, we should pair AutoML with an online monitoring and tracking system that monitors its performance and evaluates when to change the deployed model.

\noindent{\textbf{Deploying the models produced by the AutoML framework.}} To deploy an ML model in real-world systems, functionality is no longer merely dependent on accuracy but also is dependent on the inference (execution) runtime and model robustness against malicious attacks~\cite{chung2018serving, rouani2019safe}. Our current proposal does not address execution latency or safety challenges of ML models. Our domain-customized CASH optimization, however, can be modified to account for the latency and/or safety constraints of potential models. We believe co-optimizing for practical constraints such as runtime, energy consumption, or robustness against adversarial attacks is a promising future research direction.

\vspace{-3mm}
\section{Related work}
\vspace{-1mm}

The closest system to our proposed framework is Park~\cite{mao2019park}, which is an environment for experimenting with reinforcement learning algorithms for systems problems. The ML community itself has just started ramping up on AutoML research -- recent workshops and competitions have demonstrated the possibility of creating highly accurate ML models without human intervention~\cite{automlworkshop,guyon2019analysis,guyon2016brief,guyon2015design}. Similar competitions have been proposed for deep learning~\cite{liu2018autodl}.

Embedding context into AutoML frameworks helps domain experts leverage their knowledge of the problem without having to be experts in ML themselves. These context-aware AutoML frameworks produce models tailored to specific use-cases which improve their accuracy and performance (see~\S\ref{sec:motivation},~\cite{senanayake2018automorphing,orlenko2017considerations,he2018amc,mao2019park}). Such frameworks significantly reduce human effort and allow operators to take a more principled approach when using ML.

Catal et al.~\cite{catal2009systematic} studied the experience of engineering teams when building ML models for production systems and noted that an AutoML pipeline would be helpful to such teams -- especially those less experienced in using ML. We next describe several bodies of work on AutoML:

\noindent{\textbf{Hyper-parameter tuning frameworks~\cite{mendoza2016towards,feurer2018towards,falkner2018bohb,lindauer2018warmstarting,eggensperger2013towards,biedenkapp2018cave,hoos2014efficient,nisioti2018predicting}.}} Most ML models have several hyper-parameters that need to be decided before the model can be trained. Picking the right hyper-parameters for a model is crucial to achieving high accuracy and good performance. In its early forms, AutoML was mostly focused on automatically tuning the hyper-parameters of different ML models.

\noindent{\textbf{Data cleaning and pre-processing frameworks~\cite{gil2018p4ml}.}} These frameworks automatically remove missing values, do one-hot encoding of categorical values, normalize and scale data, and perform other data-cleaning operations before the dataset is input to the ML model. 

\noindent{\textbf{Feature engineering frameworks~\cite{kanter2015deep}.}} These frameworks automatically derive meaningful features from the raw data by exploiting the structure of database tables.

\noindent{\textbf{Holistic AutoML frameworks~\cite{TPoT,autosklearn,de2018automated,drori2018alphad3m,senanayake2018automorphing,rakotoarison2018automl,feurer2018practical,yang2018oboe,ghahramani2014automatic,CloudAutoML}.}} A number of AutoML frameworks target the holistic AutoML problem i.e., the CASH problem (see~\S\ref{sec:motivation}). The process involves searching through the space of hyper-parameters, data preprocessing, feature engineering, choosing the right priors and ML models, and resource allocation.

\noindent{\textbf{Context-aware AutoML~\cite{senanayake2018automorphing,orlenko2017considerations,he2018amc,CloudAutoML}.}} There have been limitted studies on context-aware AutoML frameworks. The most notable of these works is Google's Cloud AutoML platform which offers AutoML for vision, video intelligence, natural language processing, and language translation.

%1.96 *std/sqrt(n)

\newpage

\bibliographystyle{unsrt}
\bibliography{ref}

\begin{thebibliography}{10}

\bibitem{arzani2016taking}
Behnaz Arzani, Selim Ciraci, Boon~Thau Loo, Assaf Schuster, and Geoff Outhred.
\newblock Taking the blame game out of data centers operations with netpoirot.
\newblock In {\em Proceedings of the 2016 ACM SIGCOMM Conference}, pages
  440--453. ACM, 2016.

\bibitem{roy2018cloud}
Arjun Roy, Deepak Bansal, David Brumley, Harish~Kumar Chandrappa, Parag Sharma,
  Rishabh Tewari, Behnaz Arzani, and Alex~C Snoeren.
\newblock Cloud datacenter sdn monitoring: Experiences and challenges.
\newblock In {\em Proceedings of the Internet Measurement Conference 2018},
  pages 464--470. ACM, 2018.

\bibitem{roy2017passive}
Arjun Roy, Hongyi Zeng, Jasmeet Bagga, and Alex~C Snoeren.
\newblock Passive realtime datacenter fault detection and localization.
\newblock In {\em 14th $\{$USENIX$\}$ Symposium on Networked Systems Design and
  Implementation ($\{$NSDI$\}$ 17)}, pages 595--612, 2017.

\bibitem{guo2015pingmesh}
Chuanxiong Guo, Lihua Yuan, Dong Xiang, Yingnong Dang, Ray Huang, Dave Maltz,
  Zhaoyi Liu, Vin Wang, Bin Pang, Hua Chen, et~al.
\newblock Pingmesh: A large-scale system for data center network latency
  measurement and analysis.
\newblock In {\em ACM SIGCOMM Computer Communication Review}, volume~45, pages
  139--152. ACM, 2015.

\bibitem{cortez2017resource}
Eli Cortez, Anand Bonde, Alexandre Muzio, Mark Russinovich, Marcus Fontoura,
  and Ricardo Bianchini.
\newblock Resource central: Understanding and predicting workloads for improved
  resource management in large cloud platforms.
\newblock In {\em Proceedings of the 26th Symposium on Operating Systems
  Principles}, pages 153--167. ACM, 2017.

\bibitem{tammana2016simplifying}
Praveen Tammana, Rachit Agarwal, and Myungjin Lee.
\newblock Simplifying datacenter network debugging with pathdump.
\newblock In {\em 12th $\{$USENIX$\}$ Symposium on Operating Systems Design and
  Implementation ($\{$OSDI$\}$ 16)}, pages 233--248, 2016.

\bibitem{khandelwal2018confluo}
Anurag Khandelwal, Rachit Agarwal, and Ion Stoica.
\newblock Confluo: distributed monitoring and diagnosis stack for high-speed
  networks.
\newblock Technical report, Technical Report, 2018.

\bibitem{yaseen2018synchronized}
Nofel Yaseen, John Sonchack, and Vincent Liu.
\newblock Synchronized network snapshots.
\newblock In {\em Proceedings of the 2018 Conference of the ACM Special
  Interest Group on Data Communication}, pages 402--416. ACM, 2018.

\bibitem{davidblei}
David Blei.
\newblock The blessings of multiple causes.
\newblock \url{https://www.youtube.com/watch?v=Jd2nzPE7WsA}.

\bibitem{dhamdhere2007netdiagnoser}
Amogh Dhamdhere, Renata Teixeira, Constantine Dovrolis, and Christophe Diot.
\newblock Netdiagnoser: Troubleshooting network unreachabilities using
  end-to-end probes and routing data.
\newblock In {\em Proceedings of the 2007 ACM CoNEXT conference}, page~18. ACM,
  2007.

\bibitem{jalaparti2016dynamic}
Virajith Jalaparti, Ivan Bliznets, Srikanth Kandula, Brendan Lucier, and Ishai
  Menache.
\newblock Dynamic pricing and traffic engineering for timely inter-datacenter
  transfers.
\newblock In {\em Proceedings of the 2016 ACM SIGCOMM Conference}, pages
  73--86. ACM, 2016.

\bibitem{chen2014msplayer}
Yung-Chih Chen, Don Towsley, and Ramin Khalili.
\newblock Msplayer: Multi-source and multi-path leveraged youtuber.
\newblock In {\em Proceedings of the 10th ACM International on Conference on
  emerging Networking Experiments and Technologies}, pages 263--270. ACM, 2014.

\bibitem{arzani2014deconstructing}
Behnaz Arzani, Alexander Gurney, Sitian Cheng, Roch Guerin, and Boon~Thau Loo.
\newblock Deconstructing mptcp performance.
\newblock In {\em 2014 IEEE 22nd International Conference on Network
  Protocols}, pages 269--274. IEEE, 2014.

\bibitem{zhao2019minimal}
Shizhen Zhao, Rui Wang, Junlan Zhou, Joon Ong, Jeffrey~C Mogul, and Amin
  Vahdat.
\newblock Minimal rewiring: Efficient live expansion for clos data center
  networks.
\newblock In {\em Proc. USENIX NSDI}, 2019.

\bibitem{holterbach2019blink}
Thomas Holterbach, Edgar~Costa Molero, Maria Apostolaki, Alberto Dainotti,
  Stefano Vissicchio, and Laurent Vanbever.
\newblock Blink: Fast connectivity recovery entirely in the data plane.
\newblock In {\em 16th $\{$USENIX$\}$ Symposium on Networked Systems Design and
  Implementation ($\{$NSDI$\}$ 19)}, pages 161--176, 2019.

\bibitem{ghorbani2017drill}
Soudeh Ghorbani, Zibin Yang, P~Godfrey, Yashar Ganjali, and Amin Firoozshahian.
\newblock Drill: Micro load balancing for low-latency data center networks.
\newblock In {\em Proceedings of the Conference of the ACM Special Interest
  Group on Data Communication}, pages 225--238. ACM, 2017.

\bibitem{dharmapurikar2016weighted}
Sarang Dharmapurikar, Mohammadreza~Alizadeh Attar, Navindra Yadav, Ramanan
  Vaidyanathan, and Kit~Chiu Chu.
\newblock Weighted equal cost multipath routing, November~22 2016.
\newblock US Patent 9,502,111.

\bibitem{beluch2018power}
William~H Beluch, Tim Genewein, Andreas N{\"u}rnberger, and Jan~M K{\"o}hler.
\newblock The power of ensembles for active learning in image classification.
\newblock In {\em Proceedings of the IEEE Conference on Computer Vision and
  Pattern Recognition}, pages 9368--9377, 2018.

\bibitem{bottolo2010evolutionary}
Leonard Bottolo, Sylvia Richardson, et~al.
\newblock Evolutionary stochastic search for bayesian model exploration.
\newblock {\em Bayesian Analysis}, 5(3):583--618, 2010.

\bibitem{lecun2015deep}
Yann LeCun, Yoshua Bengio, and Geoffrey Hinton.
\newblock Deep learning.
\newblock {\em nature}, 521(7553):436, 2015.

\bibitem{zhang2018deepview}
Qiao Zhang, Guo Yu, Chuanxiong Guo, Yingnong Dang, Nick Swanson, Xinsheng Yang,
  Randolph Yao, Murali Chintalapati, Arvind Krishnamurthy, and Thomas Anderson.
\newblock Deepview: Virtual disk failure diagnosis and pattern detection for
  azure.
\newblock In {\em 15th $\{$USENIX$\}$ Symposium on Networked Systems Design and
  Implementation ($\{$NSDI$\}$ 18)}, pages 519--532, 2018.

\bibitem{mao2019park}
Hongzi Mao, Akshay Narayan, Parimarjan Negi, Hanrui Wang, Jiacheng Yang, Haonan
  Wang, Mehrdad Khani, Songtao He, Ravichandra Addanki, Ryan Marcus, et~al.
\newblock Park: An open platform for learning augmented computer systems.
\newblock 2019.

\bibitem{mao2017neural}
Hongzi Mao, Ravi Netravali, and Mohammad Alizadeh.
\newblock Neural adaptive video streaming with pensieve.
\newblock In {\em Proceedings of the Conference of the ACM Special Interest
  Group on Data Communication}, pages 197--210. ACM, 2017.

\bibitem{winstein2013tcp}
Keith Winstein and Hari Balakrishnan.
\newblock Tcp ex machina: Computer-generated congestion control.
\newblock 2013.

\bibitem{valadarsky2017learning}
Asaf Valadarsky, Michael Schapira, Dafna Shahaf, and Aviv Tamar.
\newblock Learning to route.
\newblock In {\em Proceedings of the 16th ACM Workshop on Hot Topics in
  Networks}, pages 185--191. ACM, 2017.

\bibitem{mao2016resource}
Hongzi Mao, Mohammad Alizadeh, Ishai Menache, and Srikanth Kandula.
\newblock Resource management with deep reinforcement learning.
\newblock In {\em Proceedings of the 15th ACM Workshop on Hot Topics in
  Networks}, pages 50--56. ACM, 2016.

\bibitem{jay2018internet}
Nathan Jay, Noga~H Rotman, P~Godfrey, Michael Schapira, and Aviv Tamar.
\newblock Internet congestion control via deep reinforcement learning.
\newblock {\em arXiv preprint arXiv:1810.03259}, 2018.

\bibitem{davemaltz}
The good, the bad, and the ugly of ml for networked systems.
\newblock
  \url{https://www.microsoft.com/en-us/research/video/the-good-the-bad-and-the-ugly-of-ml-for-networked-systems/}.

\bibitem{arzani2018007}
Behnaz Arzani, Selim Ciraci, Luiz Chamon, Yibo Zhu, Hongqiang~Harry Liu, Jitu
  Padhye, Boon~Thau Loo, and Geoff Outhred.
\newblock 007: Democratically finding the cause of packet drops.
\newblock In {\em 15th $\{$USENIX$\}$ Symposium on Networked Systems Design and
  Implementation ($\{$NSDI$\}$ 18)}, pages 419--435, 2018.

\bibitem{zhuo2017understanding}
Danyang Zhuo, Monia Ghobadi, Ratul Mahajan, Klaus-Tycho F{\"o}rster, Arvind
  Krishnamurthy, and Thomas Anderson.
\newblock Understanding and mitigating packet corruption in data center
  networks.
\newblock In {\em Proceedings of the Conference of the ACM Special Interest
  Group on Data Communication}, pages 362--375. ACM, 2017.

\bibitem{guyon2019analysis}
Isabelle Guyon, Lisheng Sun-Hosoya, Marc Boull{\'e}, Hugo~Jair Escalante,
  Sergio Escalera, Zhengying Liu, Damir Jajetic, Bisakha Ray, Mehreen Saeed,
  Mich{\`e}le Sebag, et~al.
\newblock Analysis of the automl challenge series 2015--2018.
\newblock In {\em Automated Machine Learning}, pages 177--219. Springer, 2019.

\bibitem{autosklearn}
Lars Kotthoff, Chris Thornton, Holger~H Hoos, Frank Hutter, and Kevin
  Leyton-Brown.
\newblock Auto-weka 2.0: Automatic model selection and hyperparameter
  optimization in weka.
\newblock {\em The Journal of Machine Learning Research}, 18(1):826--830, 2017.

\bibitem{TPoT}
Randal~S Olson, Nathan Bartley, Ryan~J Urbanowicz, and Jason~H Moore.
\newblock Evaluation of a tree-based pipeline optimization tool for automating
  data science.
\newblock In {\em Proceedings of the Genetic and Evolutionary Computation
  Conference 2016}, pages 485--492. ACM, 2016.

\bibitem{liu2018very}
Bin Liu.
\newblock A very brief and critical discussion on automl.
\newblock {\em arXiv preprint arXiv:1811.03822}, 2018.

\bibitem{sklearn}
scikit-learn, machine learning in python.
\newblock \url{https://scikit-learn.org/stable/index.html}.

\bibitem{guyon2016brief}
Isabelle Guyon, Imad Chaabane, Hugo~Jair Escalante, Sergio Escalera, Damir
  Jajetic, James~Robert Lloyd, N{\'u}ria Maci{\`a}, Bisakha Ray, Lukasz
  Romaszko, Mich{\`e}le Sebag, et~al.
\newblock A brief review of the chalearn automl challenge: any-time any-dataset
  learning without human intervention.
\newblock In {\em Workshop on Automatic Machine Learning}, pages 21--30, 2016.

\bibitem{bissmark2017sparse}
Johan Bissmark and Oscar W{\"a}rnling.
\newblock The sparse data problem within classification algorithms: The effect
  of sparse data on the na{\"\i}ve bayes algorithm, 2017.

\bibitem{ghahramani2014automatic}
Zoubin Ghahramani.
\newblock The automatic statistician.
\newblock 2014.

\bibitem{mukaka2012guide}
Mavuto~M Mukaka.
\newblock A guide to appropriate use of correlation coefficient in medical
  research.
\newblock {\em Malawi Medical Journal}, 24(3):69--71, 2012.

\bibitem{blumer1989learnability}
Anselm Blumer, Andrzej Ehrenfeucht, David Haussler, and Manfred~K Warmuth.
\newblock Learnability and the vapnik-chervonenkis dimension.
\newblock {\em Journal of the ACM (JACM)}, 36(4):929--965, 1989.

\bibitem{kanter2015deep}
James~Max Kanter and Kalyan Veeramachaneni.
\newblock Deep feature synthesis: Towards automating data science endeavors.
\newblock In {\em 2015 IEEE International Conference on Data Science and
  Advanced Analytics (DSAA)}, pages 1--10. IEEE, 2015.

\bibitem{CloudAutoML}
Cloud automl.
\newblock \url{https://cloud.google.com/automl/}.

\bibitem{masci2011stacked}
Jonathan Masci, Ueli Meier, Dan Cire{\c{s}}an, and J{\"u}rgen Schmidhuber.
\newblock Stacked convolutional auto-encoders for hierarchical feature
  extraction.
\newblock In {\em International Conference on Artificial Neural Networks},
  pages 52--59. Springer, 2011.

\bibitem{lakkaraju2018human}
Himabindu Lakkaraju.
\newblock {\em Human-centric Machine Learning: Enabling Machine Learning for
  High-stakes Decision-making}.
\newblock PhD thesis, Stanford University, 2018.

\bibitem{wang2019designing}
Danding Wang, Qian Yang, Ashraf Abdul, and Brian~Y Lim.
\newblock Designing theory-driven user-centric explainable ai.
\newblock In {\em Proceedings of the SIGCHI Conference on Human Factors in
  Computing Systems. CHI}, volume~19, 2019.

\bibitem{meineexplainable}
GI~Meine.
\newblock Explainable ai (ex-ai).

\bibitem{chung2018serving}
Eric Chung, Jeremy Fowers, Kalin Ovtcharov, Michael Papamichael, Adrian
  Caulfield, Todd Massengill, Ming Liu, Daniel Lo, Shlomi Alkalay, Michael
  Haselman, et~al.
\newblock Serving dnns in real time at datacenter scale with project brainwave.
\newblock {\em IEEE Micro}, 38(2):8--20, 2018.

\bibitem{rouani2019safe}
Bita~Darvish Rouani, Mohammad Samragh, Tara Javidi, and Farinaz Koushanfar.
\newblock Safe machine learning and defeating adversarial attacks.
\newblock {\em IEEE Security \& Privacy}, 17(2):31--38, 2019.

\bibitem{automlworkshop}
The third international workshop on automation in machine learning.
\newblock
  \url{https://sites.google.com/view/automl2019-workshop/home?authuser=0}.

\bibitem{guyon2015design}
Isabelle Guyon, Kristin Bennett, Gavin Cawley, Hugo~Jair Escalante, Sergio
  Escalera, Tin~Kam Ho, N{\'u}ria Macia, Bisakha Ray, Mehreen Saeed, Alexander
  Statnikov, et~al.
\newblock Design of the 2015 chalearn automl challenge.
\newblock In {\em 2015 International Joint Conference on Neural Networks
  (IJCNN)}, pages 1--8. IEEE, 2015.

\bibitem{liu2018autodl}
Zhengying Liu, Olivier Bousquet, Andr{\'e} Elisseeff, Sergio Escalera, Isabelle
  Guyon, Julio Jacques, Adrien Pavao, Danny Silver, Lisheng Sun-Hosoya,
  Sebastien Treguer, et~al.
\newblock Autodl challenge design and beta tests-towards automatic deep
  learning.
\newblock In {\em CiML workshop@ NIPS2018}, 2018.

\bibitem{senanayake2018automorphing}
Ransalu Senanayake, Anthony Tompkins, and Fabio Ramos.
\newblock Automorphing kernels for nonstationarity in mapping unstructured
  environments.
\newblock In {\em CoRL}, pages 443--455, 2018.

\bibitem{orlenko2017considerations}
Alena Orlenko, Jason~H Moore, Patryk Orzechowski, Randal~S Olson, Junmei
  Cairns, Pedro~J Caraballo, Richard~M Weinshilboum, Liewei Wang, and Matthew~K
  Breitenstein.
\newblock Considerations for automated machine learning in clinical metabolic
  profiling: altered homocysteine plasma concentration associated wtih
  metformin exposure.
\newblock In {\em Pac Symp Biocomput}, volume~23. World Scientific, 2017.

\bibitem{he2018amc}
Yihui He, Ji~Lin, Zhijian Liu, Hanrui Wang, Li-Jia Li, and Song Han.
\newblock Amc: Automl for model compression and acceleration on mobile devices.
\newblock In {\em Proceedings of the European Conference on Computer Vision
  (ECCV)}, pages 784--800, 2018.

\bibitem{catal2009systematic}
Cagatay Catal and Banu Diri.
\newblock A systematic review of software fault prediction studies.
\newblock {\em Expert systems with applications}, 36(4):7346--7354, 2009.

\bibitem{mendoza2016towards}
Hector Mendoza, Aaron Klein, Matthias Feurer, Jost~Tobias Springenberg, and
  Frank Hutter.
\newblock Towards automatically-tuned neural networks.
\newblock In {\em Workshop on Automatic Machine Learning}, pages 58--65, 2016.

\bibitem{feurer2018towards}
Matthias Feurer and Frank Hutter.
\newblock Towards further automation in automl.
\newblock In {\em ICML AutoML workshop}, 2018.

\bibitem{falkner2018bohb}
Stefan Falkner, Aaron Klein, and Frank Hutter.
\newblock Bohb: Robust and efficient hyperparameter optimization at scale.
\newblock {\em arXiv preprint arXiv:1807.01774}, 2018.

\bibitem{lindauer2018warmstarting}
Marius Lindauer and Frank Hutter.
\newblock Warmstarting of model-based algorithm configuration.
\newblock In {\em Thirty-Second AAAI Conference on Artificial Intelligence},
  2018.

\bibitem{eggensperger2013towards}
Katharina Eggensperger, Matthias Feurer, Frank Hutter, James Bergstra, Jasper
  Snoek, Holger Hoos, and Kevin Leyton-Brown.
\newblock Towards an empirical foundation for assessing bayesian optimization
  of hyperparameters.
\newblock In {\em NIPS workshop on Bayesian Optimization in Theory and
  Practice}, volume~10, page~3, 2013.

\bibitem{biedenkapp2018cave}
Andr{\'e} Biedenkapp, Joshua Marben, Marius Lindauer, and Frank Hutter.
\newblock Cave: Configuration assessment, visualization and evaluation.
\newblock In {\em International Conference on Learning and Intelligent
  Optimization}, pages 115--130. Springer, 2018.

\bibitem{hoos2014efficient}
Holger Hoos, UBC Ca, and Kevin Leyton-Brown.
\newblock An efficient approach for assessing hyperparameter importance.
\newblock In {\em International Conference on Machine Learning}, pages
  754--762, 2014.

\bibitem{nisioti2018predicting}
Eleni Nisioti, K~Chatzidimitriou, and A~Symeonidis.
\newblock Predicting hyperparameters from meta-features in binary
  classification problems.
\newblock In {\em AutoML Workshop at ICML}, 2018.

\bibitem{gil2018p4ml}
Yolanda Gil, Ke-Thia Yao, Varun Ratnakar, Daniel Garijo, Greg Ver~Steeg, Pedro
  Szekely, Rob Brekelmans, Mayank Kejriwal, Fanghao Luo, and I-Hui Huang.
\newblock P4ml: A phased performance-based pipeline planner for automated
  machine learning.
\newblock In {\em Proceedings of Machine Learning Research, ICML 2018 AutoML
  Workshop}, 2018.

\bibitem{de2018automated}
Alex~GC de~S{\'a}, Alex~A Freitas, and Gisele~L Pappa.
\newblock Automated selection and configuration of multi-label classification
  algorithms with grammar-based genetic programming.
\newblock In {\em International Conference on Parallel Problem Solving from
  Nature}, pages 308--320. Springer, 2018.

\bibitem{drori2018alphad3m}
Iddo Drori, Yamuna Krishnamurthy, Remi Rampin, Raoni de~Paula~Lourenco,
  Jorge~Piazentin Ono, Kyunghyun Cho, Claudio Silva, and Juliana Freire.
\newblock Alphad3m: Machine learning pipeline synthesis.
\newblock In {\em AutoML Workshop at ICML}, 2018.

\bibitem{rakotoarison2018automl}
Herilalaina Rakotoarison and Mich{\`e}le Sebag.
\newblock Automl with monte carlo tree search.
\newblock In {\em Workshop AutoML 2018@ ICML/IJCAI-ECAI}, 2018.

\bibitem{feurer2018practical}
Matthias Feurer, Katharina Eggensperger, Stefan Falkner, Marius Lindauer, and
  Frank Hutter.
\newblock Practical automated machine learning for the automl challenge 2018.
\newblock In {\em International Workshop on Automatic Machine Learning at
  ICML}, 2018.

\bibitem{yang2018oboe}
Chengrun Yang, Yuji Akimoto, Dae~Won Kim, and Madeleine Udell.
\newblock Oboe: Collaborative filtering for automl initialization.
\newblock {\em arXiv preprint arXiv:1808.03233}, 2018.

\end{thebibliography}
%\end{scriptsize}
\end{document}